\documentclass[english,11pt]{smfart}
\usepackage{smfthm,smfenum,amsmath,amssymb,mathrsfs,cite}
\usepackage[vcentermath]{youngtab}
\usepackage{graphicx}
\usepackage[all,poly]{xy}
\SelectTips {cm}{}

\DeclareMathOperator{\End}{End}
\DeclareMathOperator{\ch}{ch}
\DeclareMathOperator{\td}{td}

\DeclareMathOperator{\mult}{mult}
\DeclareMathOperator{\smult}{smult}

\DeclareMathOperator{\Res}{Res}

\DeclareMathOperator{\Hilb}{Hilb}

\DeclareMathOperator{\Ext}{Ext}

\DeclareMathOperator{\Li}{Li}

\DeclareMathOperator{\Pic}{Pic}

\DeclareMathOperator{\Tr}{Tr}

\numberwithin{equation}{section} 

\newcommand{\CC}{{\mathbb{C}}}

\newcommand{\RR}{\mathbb{R}}
\newcommand{\QQ}{\mathbb{Q}}
\newcommand{\ZZ}{\mathbb{Z}}

\newcommand{\PP}{\mathbb{P}}

\newcommand{\abs}[1]{\lvert#1\rvert}

\newcommand{\inv}[1]{\frac{1}{#1}}

\theoremstyle{plain}
\newtheorem{definition}[equation]{Definition}

\newtheorem{proposition}[equation]{Proposition}
\newtheorem{proposal}[equation]{Proposal}
\newtheorem{thm}[equation]{Theorem}
\newtheorem{yosou}[equation]{Conjecture}

\newtheorem{ass}[equation]{Assumption}

\theoremstyle{remark}
\newtheorem{rem}[equation]{Remark}

\author{Toshiya Kawai}
\address{Research Institute for Mathematical Sciences\\
 Kyoto University\\ Kyoto 606--8502, Japan}
\email{toshiya@kurims.kyoto-u.ac.jp}
\title{\bf String and Vortex}

\begin{document}%
\begin{abstract}
We  discuss  how  the geometry of  $D2$-$D0$  branes
  may be related to  Gromov-Witten theory of Calabi-Yau threefolds.
\end{abstract}
\thanks{Supported by Grant-in-Aid for Scientific Research \#13640017 and \#13135212.}
\maketitle
\mainmatter
%
\section{Introduction}
Topological sigma models, first put forward by Witten \cite{WittenI}, have long fascinated a number of 
theoretical physicists and mathematicians. Most remarkably, the task
of summing up worldsheet instantons is nowadays elegantly  formulated by
the theory of Gromov-Witten invariants.  Explicit computations of them are  still being actively pursued.

It goes without saying that among many possible target spaces
Calabi-Yau threefolds have played distinguished roles and are of
lasting  interest to string theorists.  Since the initial appreciation of the significance  of D-branes there has  been  the  lingering hope that the Gromov-Witten theory of
Calabi-Yau threefolds might be completely rewritten in the language of
BPS D-branes.

This contribution is intended for explaining the  picture  which, to my eye,  looks  particularly attractive in this regard. This is based on the  general philosophy:
\begin{verse}
{\em \noindent The geometry of  $D2$-$D0$ branes (and {\em not} simply $D2$-branes) provides  an alternative description of  Gromov-Witten invariants of Calabi-Yau threefolds.}
\end{verse}
 What we actually imagine is very simple and intuitive:
In analogy with the (generalized) super Kac-Moody algebras
we regard the string partition function as the inverse of ``denominator function"  and interpret  
the bound state degeneracies of $D2$-$D0$ branes
 as the ``super root multiplicities".   The Gromov-Witten potentials are then extracted from the string partition function. In particular, the variable $y$ measuring $D0$-charge is related to the genus expansion parameter $\kappa$ of Gromov-Witten theory by $y=\exp(\sqrt{-1}\kappa)$.

This sort of idea was formerly presented  in \cite{KY} when the Calabi-Yau threefold 
is elliptically fibered over a Hirzebruch surface. There,  an analogy to Borcherds products \cite{Borcherds} was pursued.  Recall that Borcherds products or their inverses arise in some cases as the denominator functions of generalized (super) Kac-Moody algebras. The first hint of the relevance of Borcherds products to Gromov-Witten theory was given by Harvey and Moore \cite{HM}. 
I will report further progress on the string partition functions of  these elliptic Calabi-Yau threefolds elsewhere \cite{K}. 

So the first purpose of this work is simply  to extend the $D2$-$D0$ picture of  \cite{KY}  to general Calabi-Yau threefolds focusing on those aspects which are believed to be independent   of any particular details of the threefolds.
 
Some time ago Gopakumar and Vafa \cite{GV1, GV2} proposed an alternative reformulation of Gromov-Witten theory of Calabi-Yau threefolds based on the space-time effective theory interpretation of Gromov-Witten potentials. This claim has been influential but at the same time very mysterious (at least to the author).
The second objective of this paper is to discuss how this proposal  of Gopakumar and Vafa  may actually reconcile with ours.

Quite recently,  
the relation between  singular instantons of $6d$ $U(1)$ gauge theory   and Gromov-Witten theory has been discussed  \cite{INOV,MNOP} in relation to the topological vertex formalism \cite{AKMV}.  This gauge theoretic approach (Donaldson-Thomas theory)  is probably a dual viewpoint of  our  $D2$-$D0$ picture in the same way point-like instantons of $4d$ $U(1)$ gauge theory describe  $D0$-branes.
 So,  as far as the ideology is concerned,  these works  seem to have some overlaps with \cite{KY} and  the present work.
Nevertheless, there is a marked difference in practice: They discuss 
the ``sum side" with a supply of explicit calculations for local toric Calabi-Yau threefolds  using the localization technique whilst we discuss the ``product side" inspired from the examples of certain elliptic Calabi-Yau threefolds and the associated Borcherds-like products \cite{KY}\cite{K}. 
It will be very interesting (and necessary!) to investigate the existence of  ``sum=product" formulas connecting both sides for a general Calabi-Yau threefold.
 In fact,   the topological vertex formalism of \cite{AKMV}  seems to allow  an  intuitive understanding in terms of   $D2$-branes and $D0$-branes.  See \S \ref{comment} for comments on this point.

\medskip

I am grateful to K.  Yoshioka for the  collaboration in \cite{KY}.
I also thank A.~Okounkov for kindly pointing out my nonsensical statement in the previous version of the manuscript and for explaining to me the marvelous  proposal of him and his collaborators.
\medskip
%
{\flushleft{\bf Notation}.}

For a rational function $f(y)$ of one variable $y$ we define $\iota_\pm f(y)\in \CC((y^{\pm 1}))$  as follows: $\iota_+ f(y)$ is  the Laurent series of $f(y)$ at $y=0$ and $\iota_- f(y)$ is that at $y=\infty$. Consider, for instance,   $\phi_h(y)=(y^{\inv{2}}-y^{-\inv{2}})^{2h-2}$ for an integer $h\ge 0$. If $h>0$ then   $\iota_+\phi_h(y)$ and $\iota_-\phi_h(y)$ coincide in  $\ZZ[y,y^{-1}]$.
However, 
\begin{equation}
  \iota_\pm \phi_0(y)=\sum_{j=1}^\infty jy^{\pm j} \in \ZZ[[y^{\pm 1}]]\,.
\end{equation}

A Calabi-Yau threefold $X$ is a complex 3-dimensional  smooth  projective variety with 
$c_1(X)=0$ and $h^{0,1}(X)=h^{0,2}(X)=0$. We assume that $X$ is polarized by some ample line bundle.

\section{Why $D2$-$D0$ rather than  $D2$ ?} \label{Why}
If one ever wishes to  connect  Gromov-Witten theory  of Calabi-Yau threefolds to some sort of  of BPS $D$-brane counting, one might think that  $D2$-branes alone are relevant since  Gromov-Witten theory is concerned with curve counting problems. However, this is too naive and even misleading. What needs to be emphasized is that in any attempt of this sort we have to incorporate the effects of $D0$-branes in addition to $D2$-branes. 

Take as an example  the case of  a resolved conifold, the total space of  $\mathscr{O}_{\PP^1}(-1)\oplus \mathscr{O}_{\PP^1}(-1)$. Since the $\PP^1$ in the resolved conifold  is rigid and the Jacobian of $\PP^1$ is just a point, the moduli space of $D2$-brane wrapping around the $\PP^1$ is also a point. If $D2$-branes were the only relevant $D$-branes, this would mean that the Gromov-Witten theory of the resolved conifold was trivial.  This is simply absurd given the result of \cite{FP}.

The origin for the necessity of $D0$-branes may be roughly as follows.
Let us suppose  that we are trying to answer the problem of counting curves in a Calabi-Yau threefold $X$. There could  be several different approaches according what we mean by {\em counting}.

In a crude approach by  $D$-branes  one may make curves ``charged" by putting  line bundles, (or more generally, rank one torsion-free sheaves) on them  and then  claim, to within signs,  the Euler-Poincar{\' e}  characteristics of the  moduli spaces of such sheaves  (regarded as torsion sheaves on  $X$) with fixed $D$-brane charges as the ``numbers of curves".  Apparently, in this naive approach we are concerned with $D2$-branes alone. 

 In Gromov-Witten theory, on the other hand,  one tries  to ``count holomorphic maps" from worldsheet connected curves to target curves in  $X$
and regard this as a good substitute of directly counting curves in $X$. However, this very substitution introduces some  well-known complications. One of them is  the so-called multi-covering effect. This is rather innocuous since we know more or less how to handle it. Another complication, which seems to be more difficult, is related to conformal invariance of the holomorphic map equation  and is known as  bubbling phenomena. 
 In order to have a nice intersection theory one must compactify the moduli space of holomorphic maps. For this we have to include degenerate contributions of bubble trees (bubbling of bubbling of \dots). 
 It  should  be precisely for this reason that  we have to modify  the naive $D2$-brane approach by including $D0$-branes 
when one attempts to rewrite  Gromov-Witten theory in terms of $D$-branes.
An intuitive picture of $D0$-branes bound to a $D2$-brane  is that of  vortices. 
So morally speaking, {\em bubble trees turn into vortices}.

\section{String partition function}
Suppose that we are given a super Kac-Moody algebra. Let $Q_+$ denote the additive semigroup generated by simple roots.  We write $\alpha>0$ iff $\alpha \in Q_+\setminus\{0\}$. Any $\alpha>0$ is either even or odd.
Consider 
\begin{equation}
\begin{split}\label{sg}
\Gamma             : &=-\sum_{\alpha>0,\, \alpha:{\rm even}} \mult(\alpha)e^{-\alpha} +\sum_{\alpha>0,\, \alpha:{\rm odd}} \mult(\alpha)e^{-\alpha}\\
&=-\sum_{\alpha>0} \smult(\alpha)e^{-\alpha}\,,
\end{split}
\end{equation}
so that $\alpha>0$ is a positive root iff $\smult(\alpha)\ne 0$.
We have 
\begin{equation}
\exp\Bigl(-\sum_{k=1}^\infty \inv {k}\psi^k(\Gamma)\Bigr) =\prod_{\alpha>0}(1-e^{-\alpha})^{-\smult(\alpha)}\,,
\end{equation}
where $\psi^k$  is the $k$-th Adams operation sending $e^{-\alpha}$ to $e^{-k\alpha}$.
The right hand side is the inverse of the denominator function in the product form.

In a nutshell, what we intend to do is to  make the analogy of this  relation   in  studying  a Calabi-Yau threefold $X$.  
We express any element of   $H_2(X,\ZZ) \oplus H_0(X,\ZZ)$  as  $(\beta,j)$ where $\beta \in H_2(X,\ZZ)$ and  $j\in \ZZ\cong H_0(X,\ZZ)$.
Denote by $N_+$  the additive semigroup of  classes of effective 1-cycles (``holomorpic curves") on  $X$. In other words, $N_+$ is the intersection of the Mori cone with $H_2(X,\ZZ)$.  We write $\beta>0$  iff $\beta \in N_+\setminus\{0\}$. Similarly, 
for  $(\beta,j)\in H_2(X,\ZZ) \oplus H_0(X,\ZZ)$  we write   $(\beta,j)>0$ iff
\begin{equation}
\bigl[ \beta>0\bigr] \quad \text{or}\quad  \bigl[\beta=0 \text{\ and\ } j>0\bigr]\,.
\end{equation}
Assume that there is a suitable $\ZZ_2$-grading so that  any $(\beta,j)>0$ is either even or odd.
Now let us  suppose in analogy with $\Gamma$ the existence of a formal sum:
\begin{equation}\label{formalsum}
  \Gamma_X=- \sum_{(\beta,j)>0} \smult(\beta,j)\mathbf{q}^\beta y^j\,, 
\end{equation}
where  $\mathbf{q}^\beta$ is a formal symbol satisfying $\mathbf{q}^\beta \mathbf{q}^{\beta'}=\mathbf{q}^{\beta+\beta'}$. 
In the proposal below,  $\smult(\beta,j)$ is in $\ZZ/2$ rather than in $\ZZ$. This is to match up with the normalization of Gromov-Witten potentials.

 We next introduce the formal truncated\footnote{Here {\it truncated} means that the ``Weyl vector"  is truncated. See Remark~\ref{remWeyl}.} free energy
\begin{equation}
  F:=-\sum_{k=1}^\infty \inv {k}\psi^k(\Gamma_X)\,,
\end{equation}
where $\psi^k$ is  again the $k$-th Adams operation sending $\mathbf{q}^\beta y^j$ to $\mathbf{q}^{k\beta}y^{kj}$.

As the inverse of the  ``denominator function"\footnote{The reason we consider  the inverse  is that we adopt   the usual rule: ``even=bosonic" \&
``odd=fermionic".} we are led to consider the following formal product
\begin{equation}
Z=\prod_{(\beta,j)>0} (1-\mathbf{q}^\beta y^j )^{- \smult(\beta,j)}\,.
\end{equation}
We might  call $Z$ as the formal truncated string partition function.
However  the expected relation ``$Z=\exp(F)$" is rather problematic since, in  the formal sum \eqref{formalsum},  $j$ runs over all integers when  $\beta>0$ so that powers of $F$ are not well-defined in general. Thus  $F$ and $Z$ in the above should be interpreted at best as motivating expressions.
In the following  we will introduce regularized versions $\tilde Z$ and $\tilde F$ related by $\tilde Z=\exp(\tilde F)$.  (These are better behaved but somewhat lose direct analogy with super Kac-Moody algebras.)

\begin{rem}
Those who are familiar with Borcherds products and their relations to surfaces, say, $K3$ surfaces will recognize that we are trying to cook up a similar story for Calabi-Yau threefolds  here.  A novel and distinct point is the  introduction of $H_0(X,\ZZ)$ in addition to $H_2(X,\ZZ)$.
 I intend to  further discuss this similarity with Borcherds products in \cite{K}.
\end{rem}

 For the above analogy to be anything useful we have to know
 $\Gamma_X$ from the geometry of  $X$.  Our basic expectation  is that $ \smult(\beta,j)$ should  
be, in some way or another,  identified  with the (super) degeneracy of bounded $D2$-$D0$ branes in $X$ with a fixed $D2$-$D0$ charge $(\beta,j)$.  
Therefore, what needs to be done is  the geometrical understanding  of $D2$-$D0$ bound systems.

\section{$D2$-$D0$ and $D2$ moduli spaces}
We first recall the gentlemen's agreement \cite{HMI} that even-dimensional $D$-branes are related to coherent sheaves and their $D$-brane charges are determined  by  Mukai vectors \cite{MukaiI,MukaiII}.  We define the $D$-brane charge $Q(\mathscr{E})$ of  a coherent sheaf $\mathscr{E}$ on $X$ by $Q(\mathscr{E})=v(\mathscr{E})\cap[X]$ where 
$v(\mathscr{E})=\ch(\mathscr{E})\sqrt{\td(X)}$ is the Mukai vector of $\mathscr{E}$.
We express $Q(\mathscr{E})$ in the form $(Q_6,Q_4,Q_2,Q_0)$ where
 $Q_{2i}\in H_{2i}(X,\QQ)$. 
Note that if the $D$-brane charge is of the form  $(0,0,Q_2,Q_0)$ then $(Q_2,Q_0)$ is actually  in $H_2(X,\ZZ)\oplus H_0(X,\ZZ)$.  (There is no Witten effect.)
For instance, suppose that $i:C\hookrightarrow X$ is  a smooth irreducible curve of genus $g$ and $L\to C$ is a  line bundle of degree $d$, then $Q(i_*L)=(0,0,[C], \chi(C,L))=(0,0,[C],d+1-g)$. This is the $D$-brane charge of a $D2$-brane singly wrapping around $C$ bound with $d$ $D0$-branes.  In this paper any $D2$-brane is always singly wrapping by allowing non-reduced curves.

If  the  $D$-brane charge (or equivalently the Mukai vector) is fixed, the Hilbert polynomial is also fixed since $X$ is assumed to be polarized. 
So it makes sense
to consider the moduli space $\mathcal{M}(Q)$ of semi-stable coherent sheaves on $X$ with a fixed $D$-brane charge $Q$. For simplicity let us suppose that $\mathcal{M}(Q)$ consists only of stable sheaves. One well-known  fact about $\mathcal{M}(Q)$ is that its expected dimension $\sum_{i=0}^3(-1)^{i+1}\dim \Ext_0^i(\mathscr{E},\mathscr{E})$  vanishes because of the Serre duality where $\mathscr{E}\in \mathcal{M}(Q)$ and $\Ext_0^i$ is the trace free part of $\Ext^i$. 
By stability, $\Ext_0^0(\mathscr{E},\mathscr{E})\cong \{0\}\cong \Ext_0^3(\mathscr{E},\mathscr{E})$ and by $h^{0,1}(X)=h^{0,2}(X)=0$, we have $\Ext^i(\mathscr{E},\mathscr{E})=\Ext_0^i(\mathscr{E},\mathscr{E})$ $(i=1,2)$. The zero-dimensional virtual moduli cycle  was constructed in \cite{Thomas}.  Its degree or virtual length $\lambda(\mathcal{M}(Q))\in \ZZ$ serves as  the ``number of sheaves".  
 The Zariski tangent space of $\mathcal{M}(Q)$ at $\mathscr{E}$ is given by $\Ext^1(\mathscr{E},\mathscr{E})$ and the obstruction space is $\Ext^2(\mathscr{E},\mathscr{E})$. Hence the Serre duality tells us that if $\mathcal{M}(Q)$ is smooth, the obstruction sheaf is the cotangent bundle of  $\mathcal{M}(Q)$ and $\lambda(\mathcal{M}(Q))=(-1)^{\dim\mathcal{M}(Q)}\chi(\mathcal{M}(Q))$.

One can also consider  the Hilbert scheme $\Hilb(Q)$ with a fixed $D$-brane charge $Q$.
As usual, we have $\Hilb(Q)=\mathcal{M}(Q(\mathscr{O}_X)-Q)$ by considering ideal sheaves.
The  moduli space of possibly disconnected 
$D2$-branes and $D0$-branes, which we call the total $D2$-$D0$ moduli space,  is given by 
\begin{equation}\label{totalhilb}
\coprod_{(Q_2,Q_0)}\Hilb(0,0,Q_2,Q_0)\,,
\end{equation}
where  $(Q_2,Q_0)$ runs  over all possible values. 
At  first sight \eqref{totalhilb} seems not to deserve its name. For instance, if 
an isolated  $D2$-brane is singly wrapping around a smooth curve $C\subset X$, there should exist degrees of freedom in how it wraps, namely  the Jacobian $J(C)$.
On the other hand in the Hilbert scheme $C$ is allowed to carry only $\mathscr{O}_C$.
A possible resolution of this puzzle may be as follows. First of all if a $D2$-brane wrapping singly around $C$ is bounded to several $D0$-branes, the moduli space of this bound system is given by a symmetric product of $C$ (as we recall later). So this part of the moduli space is directly related to the Hilbert scheme \eqref{totalhilb}. 
The pure $D2$-brane moduli space $J(C)$ is actually related to the 
  symmetric products of $C$ through the Abel-Jacobi maps. 
 So, once we start to consider both $D2$-branes and $D0$-branes simultaneously, we should forget about $J(C)$ and only consider the symmetric products of  $C$ as fundamental in order  not to overcount the degrees of freedom.  A more explanation is given below about how pure $D2$-brane moduli spaces are related to the moduli spaces of bounded $D2$-$D0$ branes. (This is not unrelated to the upcoming interpretation of the Gopakumar-Vafa proposal.)

A  relation between $\lambda(\mathcal{M}(Q))$ and Gromov-Witten  invariants was already hinted in \cite{Thomas}.  A  connection to the Gopakumar-Vafa invariant was  conjectured in \cite{HST}.
Recently,   a  striking connection of $\lambda(\Hilb(Q))$ to Gromov-Witten theory has been proposed  in  \cite{MNOP}.

We expect that  $\Gamma_X$ is related to  \eqref{totalhilb} but exactly describing this relation seems to be a very difficult problem at the moment.  Only a limited attempt is given below.
Let us call $\beta\in N_+$ {\em of simple class\it} if its arithmetic genus $g(\beta)$ is non-negative and for each  integer $d\ge 0$ there exist a suitable $D2$-$D0$ bound moduli space $\mathfrak{M}_{\beta,d}$ and a $D2$ moduli space $\mathfrak{N}_{\beta,d}$ whose properties we spell out in the following.
 We assume that $\beta=0$ is of simple class and set  $g(0)=0$.

 The $D2$-$D0$ bound moduli space $\mathfrak{M}_{\beta,d}$ describes a $D2$-brane  wrapping around a curve of class $\beta$ bound to $d$ $D0$-branes. In particular we expect an intimate connection between
 $\mathfrak{M}_{\beta,d}$ and  $\Hilb(0,0,\beta,d+1-g(\beta))$. We assume that $\mathfrak{M}_{\beta,d}$ is smooth and connected.
As we will see shortly, it seems natural to have  
 \begin{equation}\label{pureD0}
{\mathfrak{M}_{0,d}}=X\times \PP^d\,.
\end{equation}
We set $S_\beta:=\mathfrak{M}_{\beta,0}$ for convenience. This is the moduli space of the {\it supports\/} of  $D2$-branes. Note that $S_0=X$ by \eqref{pureD0}.
There should exist  a  morphism $\mathfrak{M}_{\beta,d}\to S_\beta$ and
\begin{equation}
\dim\mathfrak{M}_{\beta,d}=S_\beta+d\,.
\end{equation}
 Moreover, there will be the universal family $\mathcal{C}_\beta\to S_\beta$
which is a flat family of curves of class $\beta$ and of arithmetic genus $g(\beta)$.
In particular we expect
\begin{equation}\label{C_0S_0}
\mathcal{C}_0=X\times \PP^1\,,
\end{equation}
where the morphism $\mathcal{C}_0\to S_0$ is  the projection.
Intuitively speaking, $\mathcal{C}_{\beta}\to S_\beta$  is the family of curves around which $D2$-branes are singly wrapping. 
In favorable situations it is tempting to identify ${\mathfrak{M}_{\beta,d}}$ with
 the relative Hilbert scheme of points on curves 
\begin{equation}
\mathcal{C}_\beta^{[d]}:=\Hilb^d_{\mathcal{C}_\beta/ S_\beta}\,.
\end{equation}
Since we assumed that $\mathfrak{M}_{\beta,d}$ is smooth, our concern is limited to a smooth $\mathcal{C}_\beta^{[d]}$.

Let us explain why  \eqref{pureD0} and \eqref{C_0S_0} seem natural. We regard ${\mathfrak{M}_{0,d}}$ as the moduli space of the bound system of a $D2$-brane and $d$ $D0$-branes in the limit where the support curve of the  $D2$-brane   is  shrinking to a point. What does the curve look like?  Since we do not expect any $D2$-brane degree of freedom in the end, the Jacobian of the curve must be a point. So the curve will be  a $\PP^1$. With the vortex interpretation given in \cite{KY} and to be recalled later, the moduli space of the $D2$-brane wrapping around $\PP^1$ and $d$ $D0$-branes sticked to it should be given by the $d$-th symmetric product of $\PP^1$ or $\PP^d$. Therefore if we take  into account the location of the shrinking $\PP^1$ in $X$, we are led to \eqref {pureD0} and  \eqref{C_0S_0}.

Let us turn to the properties of the 
$D2$  moduli space. We assume that  $\mathfrak{N}_{\beta,d}$ is smooth and connected.
There should exist a natural morphism 
 $\mathfrak{N}_{\beta,d}\to S_\beta$. Intuitively, $\mathfrak{N}_{\beta,d}$  is the moduli space of $D2$-branes singly wrapping around fibers of $\mathcal{C}_\beta\to S_\beta$. 
We expect an intimate connection between   $\mathfrak{N}_{\beta,d}$ and  $\mathcal{M}(0,0,\beta,d+1-g(\beta))$.
We furthermore expect that $\mathfrak{N}_{\beta,d}$ is independent of $d$ and is isomorphic to $\mathfrak{N}_\beta:=\mathfrak{N}_{\beta,0}$.
We should have  $\mathfrak{N}_0=X$. Hence $\mathfrak{N}_0\to S_0$ is the identity map. 
We require the existence of  a commutative diagram
\begin{equation}\label{MNSdiagram}
\xymatrix{&\mathfrak{M}_{\beta,d}\ar[dl]\ar[dr]&\\
\mathfrak{N}_\beta\ar[rr]&&S_\beta}
\end{equation}

  For $\beta=0$ we have \eqref{MNSdiagram} by obvious morphisms. For $d=0$, $\mathfrak{M}_{\beta,0}=S_\beta \to \mathfrak{N}_\beta$ may be viewed as taking a section.

If $\mathfrak{M}_{\beta,d}=\mathcal{C}_\beta^{[d]}$, one may take
\begin{equation}
\mathfrak{N}_{\beta,d}=\overline{\Pic}^d_{\mathcal{C}_\beta/ S_\beta}\,, \qquad
\mathfrak{N}_{\beta}=\overline{\Pic}^0_{\mathcal{C}_\beta/ S_\beta}=:\bar{\mathcal{J}}_\beta\,,
\end{equation}
where $\overline{\Pic}^d_{\mathcal{C}_\beta/ S_\beta}$ is the relative compactified Picard scheme and $\bar{\mathcal{J}}_\beta$ is the relative compactified Jacobian.
Then the diagram \eqref{MNSdiagram} is replaced by
\begin{equation}\label{CJSdiagram}
\xymatrix{&\mathcal{C}^{[d]}_\beta\ar[dl]\ar[dr]&\\
\bar{\mathcal{J}}_\beta\ar[rr]&&S_\beta}
\end{equation}
The morphism $\mathcal{C}_\beta^{[d]}\to \bar{\mathcal{J}}_\beta$ is  just the Abel-Jacobi map.

We should have
\begin{equation}
\dim \mathfrak{N}_\beta=\dim S_\beta+g(\beta)\,.
\end{equation}

This ends  the list of what  we demand for $\beta$ to be of simple class.

We now  turn to our proposal on the structure of $\Gamma_X$:
\begin{proposal}
By taking into account  charge conjugation symmetry,   
we have
\begin{equation} 
\begin{split}
 2\Gamma_X=&\sum_{d=0}^\infty \nu(0,d)y^{d+1}\\
&\ +\sum_{\beta>0} \sum_{d=0}^\infty \nu(\beta,d)\left(y^{d+1-g(\beta)}+y^{-(d+1-g(\beta))}\right)\,\mathbf{q}^\beta\,,
\end{split}
\end{equation}
where $\nu(\beta,d)\in \ZZ$ and if $\beta$ is of simple class
$\nu(\beta,d)=\epsilon(\beta)\chi(\mathfrak{M}_{\beta,d})$ with
\begin{equation}\label{sign}
\epsilon(\beta)=(-1)^{\dim S_\beta+g(\beta)}=(-1)^{\dim \mathfrak{N}_\beta}\,.
\end{equation}
\end{proposal}
This looks like  the most  natural  extension of  what  we conceived in \cite{KY}.  The reason why we choose the particular sign factor \eqref{sign} is explained below by comparison with Gromov-Witten theory. 
Note that even if $\mathfrak{M}_{\beta,d}$ is non-empty, $\chi(\mathfrak{M}_{\beta,d})$ can vanish.  In our analogy with super Kac-Moody algebras this is  related  to the vanishing of $\smult(\beta,j)$ when $(\beta,j)$ is not a   ``positive root". 

At the moment we do not know how exactly $\nu(\beta,d)$ should be described geometrically when $\beta$ is other than  of simple class.

\section{Gromov-Witten potentials}

To make contact with Gromov-Witten theory we further postulate
\begin{ass}
For each $\beta\in N_+$ there exists a rational function
$h_\beta(y)$  with  inversion symmetry $h_\beta(y)=h_\beta(y^{-1})$ such that
\begin{equation}
  \sum_{d=0}^\infty \nu(\beta,d)y^{\pm(d+1-g(\beta))}=\iota_\pm   h_\beta(y)\,.
\end{equation}
If $\beta$ is of simple class,  we introduce $f_\beta(y)$ by $h_\beta(y)=\epsilon(\beta)f_\beta(y)$, namely,
\begin{equation}\label{fund}
  \sum_{d=0}^\infty \chi( \mathfrak{M}_{\beta,d}   )y^{\pm(d+1-g(\beta))}=\iota_\pm   f_\beta(y)\,.
\end{equation}
Moreover we have  the  expansion of the form
\begin{equation}\label{expansionoffbeta}
h_\beta(y)=-\sum_{g=0}^\infty r^g_\beta\, \kappa^{2g-2}\,,
\end{equation}
where  $y=\exp(\sqrt{-1}\kappa)$ and $r^g_\beta\in \QQ$.
\end{ass}
One then observes that
\begin{equation}\label{Gammatilde}
 2\Gamma_X=\iota_+h_0(y)+\sum_\pm \iota_\pm
\sum_{\beta>0}h_\beta(y)\, \mathbf{q}^\beta\,.
\end{equation} 

So far $\mathbf{q}^\beta$ has been a formal symbol, but in Gromov-Witten theory we should like to set
$\mathbf{q}^\beta=\exp(\omega \cap \beta)$ where $\omega$ is the complexified K{\" a}hler form of $X$. 
Therefore, motivated by \eqref{Gammatilde}, we introduce
\begin{equation}
  \widetilde{\Gamma}_X=\inv{2}h_0(y)+
\sum_{\beta>0}h_\beta(y)\, \mathbf{q}^\beta\,,
\end{equation}
and  interpret this as a power series expansion in $q_i:=\exp(\omega\cap \beta_i)$ $(i=1,\dots,h^{1,1}(X))$.
Here  $\beta_1,\dots,\beta_{h^{1,1}(X)}\in H_2(X,\ZZ)$ are generators of the Mori cone.
In addition one may introduce the truncated free energy  by
\begin{equation}
\tilde F:=-\sum_{k=1}^\infty\inv{k}\psi^k(\widetilde{\Gamma}_X)\,,
\end{equation}
and the truncated string partition function $\tilde Z:=\exp(\tilde F)$. Then, if $\abs{\mathbf{q}^\beta y^{d+1-g(\beta)}}\ll 1$ for all $\beta \ge 0$ and $d \ge 0$, we have an infinite product representation :
\begin{equation}\label{infprod}
\tilde Z=\prod_{\beta\ge 0}\prod_{d=0}^\infty(1-\mathbf{q}^\beta y^{d+1-g(\beta)})^{\mu(\beta,d)}\,,
\end{equation}
where
\begin{equation}
\mu(\beta,d)=\begin{cases} \inv{2}\nu(0,d),&\beta=0,\\ 
\nu(\beta,d),&\beta>0.
\end{cases}
\end{equation}
By \eqref{pureD0} we have $\chi(\mathfrak{M}_{0,d})=(d+1)\chi(X)$. Hence we have
 $\mu(\beta,d)\in \ZZ$ because of  $\chi(X)\in 2\ZZ$.

To discuss the relation  to Gromov-Witten theory we want to set $y=\exp(\sqrt{-1}\kappa)$ with
$\kappa\in \RR$ and $\abs{\kappa}$ very small and interpret $\kappa$ as the genus expansion parameter.  As above we always assume  $\abs{\mathbf{q}^\beta}\ll1$ for $\beta>0$ in the following\footnote{However, as is often the case in Gromov-Witten theory we neglect the issue of convergence.}.
For such expansions in $\kappa$ to be possible, the truncated  free energy needs a bit of regularization:
\begin{equation}
\tilde F^s:=-\sum_{k=1}^\infty\inv{k^s}\psi^k(\widetilde{\Gamma}_X)\,\quad(s\in\CC,\ \Re(s)>1)\,.
\end{equation}
Let us  introduce $F_g^s$ via the expansion
$\tilde F^s=\sum_{g=0}^\infty \kappa^{2g-2}F_g^s$  and define
\begin{equation}
F_g:=\lim_{s\to 1}F_g^s\,,\quad(g\ne 1),\qquad  F_1:=\lim_{s\to 1}\left[F_1^s-\inv{2}r_0^1\left(\inv{s-1}+\gamma_{\rm em}\right)\right],
\end{equation}
where $\gamma_{\rm em}$ is the Euler-Mascheroni constant.
\begin{yosou}
The Gromov-Witten potentials ${\bf F}_g$ $(g\ge 0)$ of $X$ are given by
\begin{equation}
{\bf F}_0=F_0^{\rm cl}+F_0, \quad  {\bf F}_1=F_1^{\rm cl}+F_1, \quad {\bf F}_g=F_g\ \  (g>1)\,,
\end{equation}
where 
\begin{equation}
F_0^{\rm cl}=\inv{3!}\int_X\omega^3\,,\qquad F_1^{\rm cl}=-\inv{24}\int_X c_2(X)\, \omega\,.
\end{equation}
\end{yosou}
\begin{rem}\label{remWeyl}
In seeking analogy with super Kac-Moody algebras $\kappa^{-2}F_0^{\rm cl}+F_1^{\rm cl}$ should be interpreted as (the negative of) the ``Weyl vector". This part is quite delicate and important.
\end{rem}

In fact by using  \eqref{expansionoffbeta}   we find that
\begin{equation}
  F_g=\frac{1}{2}r_0^{g }\,\zeta^*(3-2g)+\sum_{\beta>0}r^g_\beta\, \Li_{3-2g}(\mathbf{q}^\beta)\,,
\end{equation}
where $\zeta^*(s):=\zeta(s)$ if $s\ne 1$ and $\zeta^*(1):=0$.  This is an expected form.

Let us give some evidence for our choice \eqref{sign}. Note first that $\epsilon(0)=(-1)^3=-1$ and 
\begin{equation}
f_0(y)=\frac{\chi(X)}{(y^{\inv{2}}-y^{-\inv{2}})^{2}}\,.
\end{equation}
This enables us to  calculate  $r_0^{g}$ explicitly.
For $g=0$,  we have $\frac{1}{2}r_0^{0 }\,\zeta(3)=-\inv{2}\chi(X)\zeta(3)$.
This is familiar to us.
  As for $g>1$,  the constant map contribution is correctly reproduced:
\begin{equation}\label{constantmap}
\frac{1}{2} r_0^{g}\,\zeta(3-2g)= \frac{1}{2}\chi(X)(-1)^g\int_{\overline{\mathcal{M}}_{g,0}}(\lambda_{g-1})^3\,,\qquad (g>1)\,,
\end{equation}
where $\overline{\mathcal{M}}_{g,0}$ is the Deligne-Mumford moduli stack of stable curves of arithmetic genus $g$ without marked points and $\lambda_{g-1}$ is the $(g-1)$-th Chern class of the Hodge bundle  over $\overline{\mathcal{M}}_{g,0}$. 
The relation \eqref{constantmap} was conjectured physically in  \cite{GV1} from $M$-theory interpretation and in \cite{MM} from duality to heterotic string. A mathematical proof of \eqref{constantmap} was given in \cite{FP}. An argument here from a purely $D2$-$D0$ point of view is new.

Suppose that $\beta$ is of simple class and that  both $\mathcal{C}_\beta$ and $S_\beta$ are smooth. Assume that {\it all} the fibers of $\mathcal{C}_\beta\to S_\beta$ are  smooth irreducible curves of genus $g(\beta)$. Then  the vortex interpretation explained below shows that
\begin{equation}\label{smooth}
f_\beta(y)=\chi(S_\beta)(y^{\inv{2}}-y^{-\inv{2}})^{2g(\beta)-2}\,.
\end{equation}
If  $\epsilon(\beta)=(-1)^{\dim S_\beta+g(\beta)}$  we see that  $r^{g(\beta)}_\beta=(-1)^{\dim S_\beta}\chi(S_\beta)=\int_{S_\beta}c_{\rm top}(T_{S_\beta}^\vee)$. This is consistent with   Gromov-Witten theory.

\section{Comments on the ``sum side"}\label{comment}
Not infrequently  we encounter non-trivial formulas of the form
\begin{equation}\label{denom}
\text{sum=product.}
\end{equation}
So far we have been discussing  the product side.
It is natural to ask what the sum side looks like.    
Obviously, the sum should be taken over potentially all the $D2$-$D0$ states. So the total $D2$-$D0$ moduli space \eqref{totalhilb} will be relevant.
Quite recently \cite{INOV,MNOP}, in an attempt to understand the formalism of the topological vertex for local Calabi-Yau threefolds, the connection between  singular instantons of $6d$ $U(1)$ gauge theory  and  Gromov-Witten theory  has been studied.  
This is expected to be another description of our  $D2$-$D0$ picture studied on the sum side. In particular, \cite{MNOP} treats local Calabi-Yau threefolds  and evaluates, by the localization technique,  the  generating function of $\lambda(\mathcal{M}(Q))$ for the moduli spaces of ideal sheaves $\mathcal{M}(Q)=\Hilb(Q(\mathscr{O}_X)-Q)$. This is a very explicit and remarkable calculation.

  In fact the topological vertex formalism \cite{AKMV,ORV} fits quite nicely into the picture of $D2$-$D0$ branes. 
For a  local toric Calabi-Yau threefold, the relevant sum is  reduced by localization to the one over  the  torus-fixed configurations of $D2$-$D0$ branes.  In the approach of \cite{AKMV}  one writes the diagram consisting of edges and trivalent vertices. Each internal edge corresponds to a $(\CC^\times)^2$-fixed  $\PP^1$.
Each vertex corresponds to a $(\CC^\times)^3$-fixed point and is the north or south poles of the $\PP^1$'s of the internal edges emanating from it. It is clear that the  torus-fixed configurations of $D2$-$D0$ branes are such that $D2$-branes are localized to the  $(\CC^\times)^2$-fixed  $\PP^1$'s (the internal edges) and $D0$-branes are localized to the $(\CC^\times)^3$-fixed points (the vertices).
The way several $D2$-branes are localized to the internal edge $\PP^1$'s must be treated scheme-theoretically  and described by assigning  a $2d$ Young diagram to each edge. 
Similarly, the way $D0$-branes accumulate to  the vertex points must also be treated scheme-theoretically and described by assigning a $3d$ Young diagram to each vertex.  However 
the configurations of $D0$-branes and those of  $D2$-branes are not independent. How these are actually linked is specified by the rule of the topological vertex.

It should be mentioned that   having a nice sum expression does not a priori guarantee the connection to Gromov-Witten theory. For such to exist  ``sum=product" will probably have to hold. 
For super (generalized) Kac-Moody algebras, this problem is related to  finding the set of positive roots among $Q_+$, which in general is a subtle and difficult task.

Obviously, if our ``product'' proposal and the ``sum" proposal (GW/DT correspondence) of \cite{MNOP}  are consistent, $\{\nu(\beta,d)\}$ and $\{\lambda(\Hilb(0,0,Q_2,Q_0)\}$ must be related. This looks plausible since both are concerned with counting $D2$ and $D0$ branes in $X$.
However, the precise geometrical characterization of $\{\nu(\beta,d)\}$ is yet to be found.

\section{Comparison with Gopakumar-Vafa proposal}

In  \cite{GV2}, Gopakumar and Vafa  asserted that the $D2$-brane moduli spaces of  a Calabi-Yau threefold $X$ have some characteristic properties.  We considered the $D2$ moduli space $\mathfrak{N}_\beta$ when $\beta$ is of simple class and assumed that 
$\mathfrak{N}_\beta$ is smooth. To recall the Gopakumar-Vafa proposal, let us temporarily lift the smoothness condition of   $\mathfrak{N}_\beta$. In this slightly more general context, we cast their claim taking into account the suggestion of \cite{HST} as follows:

\medskip
\begin{em} 
Let $IH^*(\mathfrak{N}_\beta)$ denote the intersection cohomologies of $\mathfrak{N}_\beta$.
There exists a  representation $\rho:sl(2)\oplus sl(2)\to \End(IH^*(\mathfrak{N}_\beta))$  in such a way that the restricted  representations $\rho_L,\rho_R,\rho_D:sl(2)\to\End(IH^*(\mathfrak{N}_\beta)) $  associated  respectively with
the first, the second and the diagonal $sl(2)$ subalgebras of $sl(2)\oplus sl(2)$ have the following interpretations: $\rho_L$ resp.~$\rho_R$ corresponds to  the Lefschetz action   in  the ``fiber  resp.~base direction" of the cohomologies of $\mathfrak{N}_\beta\to S_\beta$
while $\rho_D$ corresponds to the usual Lefschetz action. Hence one can introduce
\begin{equation}
 \Lambda_\beta(y):=\Tr_{IH^*(\mathfrak{N}_\beta)} (-1)^{H_D} y^{H_L}\in \ZZ[y,y^{-1}]\,,
 \end{equation}
where $H_L$ and $H_D$  are respectively the images by $\rho_L$ and $\rho_D$ of the Cartan generator of $sl(2)$. (The eigenvalues of  $H_L$ and $H_D$ are twice spins.)
Note the obvious symmetry $\Lambda_\beta(y^{-1})=\Lambda_\beta(y)$.
\end{em}
\medskip

 According to \cite{HST}, this claim  is true
if $\mathfrak{N}_\beta\to S_\beta$ is a
 projective morphism between two normal projective varieties. 

In the rest of the paper we only consider the cases when $\beta$'s are of simple class.
Hence $\mathfrak{M}_{\beta,d}$ and $\mathfrak{N}_\beta$ do exist and they are smooth and connected.  When dealing with $\mathfrak{N}_\beta$ we do not need $IH^*(-)$ and $H^*(-)$ will suffice.

Gopakumar and Vafa  argued  that $\Lambda_\beta(y)$'s  determine the Gromov-Witten potentials of $X$.  In particular, considering certain  combinations of   irreducible representations of $sl(2)$ to be  fundamental, they expanded $\Lambda_\beta(y)$ as 
\begin{equation}\label{GVexpansion}
\epsilon(\beta)\Lambda_\beta(y)=\sum_{h=0}^{g(\beta)} N_\beta^h\,(y^{\inv{2}}-y^{-\inv{2}})^{2h}\,.
\end{equation}
Then they considered that the {integers} $N_\beta^h$ are alternative  fundamental invariants\footnote{Their definition of the invariants differ in signs with $N_\beta^h$ here.}. 
 Since $H_D$ acts as the multiplication of $\dim\mathfrak{N}_\beta-k$ on
 $H^k(\mathfrak{N}_\beta)$,
we have  $\epsilon(\beta)\Lambda_\beta(1)=\chi(\mathfrak{N}_\beta)=N_\beta^0$.

This claim of Gopakumar and Vafa may sound somewhat at odds with our earlier statement: {\em Considering $D2$-branes alone is insufficient}.  Actually they  mapped the problem  into $M$-theory  and used a physical  interpretation of space-time effective theory typically using Schwinger-like computations. This is  the reason why $D0$-branes secretly sneak in  when interpreted in terms  of type IIA theory.

The  consistency of our proposal with that of Gopakumar and Vafa implies
\begin{yosou} 
\begin{equation}\label{K-GV}
f_\beta(y)=\frac{\epsilon(\beta)\Lambda_\beta(y)}{(y^{\inv{2}}-y^{-\inv{2}})^2}\,.
\end{equation}
\end{yosou}
\begin{rem}
Since $\mathfrak{N}_0=S_0=X$ and the ``fiber direction" is void, we have $\Lambda_0(y)=-\chi(X)$.
Consequently \eqref{K-GV} is true for $\beta=0$.
\end{rem}

Despite its innocent looking  the geometrical  implication of  \eqref{K-GV}  is rather non-trivial since  the interpretations are quite different on both sides of the equation.

\section{Inversion relations}

Before discussing the validity of \eqref{K-GV}
let us pause to study purely algebraic consequences of \eqref{K-GV}.
We expand $\iota_\pm f_\beta(y)$ as \eqref{fund} and $\Lambda_\beta(y)$  as \eqref{GVexpansion}. If we set
\begin{equation}
g=g(\beta), \qquad e_d=\chi(\mathfrak{M}_{\beta,d}),\qquad m_i=N_\beta^{g(\beta)-i}\,,
\end{equation}
 then the following  proposition gives the explicit relations between
$\chi(\mathfrak{M}_{\beta,d})$ and $N_\beta^h$. 
Note  that since $e_0=\chi(S_\beta)$ by our definition, the relation $e_0=m_0$
below  implies $N_\beta^{g(\beta)}=\chi(S_\beta)$.

\begin{proposition}\label{invprop}
Let $g \ge 0$ be an integer. Suppose that sequences of numbers $\{e_d\}_{d=0}^\infty$ and $\{m_i\}_{i=0}^g$ are related by
\begin{equation} \label{AJ}
  \sum_{d=0}^\infty e_d\,  y^{\pm(d+1-g)}=\iota_\pm \sum_{i=0}^g m_i\,  (y^{\inv{2}}-y^{-\inv{2}})^{2(g-i)-2}\,.
\end{equation}
\flushleft{\rm (A)}\ One can express $\{e_d\}_{d=0}^\infty$ in terms of $\{m_i\}_{i=0}^g$ as:
\begin{align}
&\ast\ g=0:&&&\nonumber\\
    && e_d&=(d+1)m_0\,, &(d \ge 0)\,,\\
&\ast\ g=1:& &&\nonumber\\
&&      e_0&=m_0\,,      & \\
               & & e_d&=d\, m_1\,,       & (d \ge 1 )\,,\\
&\ast\ g\ge 2:&&&\nonumber\\
&& e_d&=\sum_{i=0}^d\binom{d+i +1-2g}{d-i}m_{i}\,, 
& ( g-1 \ge d\ge 0)\,, \\
&&e_d&=\sum_{i=0}^{2g-2-d}\binom{d+i +1-2g}{d-i}m_{i}
& \\
&&&\hspace{2.5cm}+(d+1-g)m_g\,,&(2g-2\ge d\ge g)\,, \nonumber\\
&&e_d&=(d+1-g)m_g\,, 
& (d\ge 2g-1)\label{projectivebundle}\,.
\end{align}

\flushleft{{\rm (B)}}\ Conversely, $\{m_i\}_{i=0}^g$ can be expressed in terms of  $\{e_d\}_{d=0}^g$ as:
\begin{align}
&\ast\  g=0:&&&\nonumber\\
&&   m_0&=e_0\,,&\\
&\ast\  g=1:&&&\nonumber\\
&&   m_0&=e_0\,,\quad
  m_1=e_1\,,&\\
&\ast\  g\ge 2:&&&\nonumber\\
&&m_0&=e_0\,,&\\
  &&m_{i}&=e_i+\sum_{d=0}^{i-1}\frac{2g-2d -2}{i-d}\binom{2g-i-d-3}{i-d-1}e_d\,,& \label{invrel}\\
  &&&\hspace{6cm}( g-1 \ge i \ge 1)\,,& \nonumber\\
  &&m_g&=e_g-e_{g-2} \label{m_g}\,.&
\end{align}

\flushleft{\rm (C)}\
If  $\{e_d\}_{d=0}^\infty$  are integers, so are $\{m_i\}_{i=0}^g$  and vice versa.
\end{proposition}
\begin{proof}  \eqref{AJ} implies the following Taylor expansion at $y=0$:
\begin{equation}\label{tmpeq1}
    \sum_{d=0}^\infty e_d\,  y^d=\sum_{i=0}^g m_{i}\, y^i (1-y)^{2(g-i)-2}\,.
  \end{equation}
It suffices to analyze this equation.
{\flushleft \rm (A)} 
This is a consequence of straightforward binomial expansions of the right hand side of \eqref{tmpeq1}. 
\flushleft{\rm (B)} Since the cases $g=0$ and $g=1$ are obvious, we assume $g\ge 2$. Notice  that \eqref{tmpeq1} is equivalent to
\begin{equation}
   \frac{\sum_{d=0}^\infty e_d\, y^d}{(1-y)^{2g-2}}=
\sum_{i=0}^g m_{i}\,\left(\frac{y}{(1-y)^2}\right)^i\,.
\end{equation}
Then, by the Lagrange inversion formula,  one finds
\begin{equation}
  \begin{split}
    m_{i}&=\Res_{y=0}\left[
\frac{\sum_{d=0}^\infty e_d\, y^d}{(1-y)^{2g-2}}
\left(\frac{y}{(1-y)^2}\right)'\left(\frac{y}{(1-y)^2}\right)^{-i-1}\right]\\
           &= \Res_{y=0}\left[\frac{1+y}{(1-y)^{2(g-i)-1}}\sum_{d=0}^\infty e_d\, y^{d-i-1}\right],
  \end{split}
\end{equation}
where $'$ represents the derivative with respect to $y$.
To evaluate this residue it is better to proceed case by case:\\

\flushleft{ $\ast$} Suppose that   $2(g-i)-1\le 0$.
 Then we  actually have to have $g=i$ and
\begin{equation}
  m_g=\Res_{y=0}\left[(1-y^2)\sum_{d=0}^\infty e_d\, y^{d-g-1}\right].
\end{equation}
It is straightforward to evaluate this residue.   The result is given by \eqref{m_g}. \\
\flushleft{ $\ast$} Suppose instead that   $2(g-i)-1> 0$.
 One uses the formula\begin{equation}
  \inv{(1-y)^k}=\sum_{n=0}^\infty \binom{k+n-1}{n} y^n\,,
\end{equation}
(valid for a positive integer $k$) to obtain
\begin{equation}
  m_{i}=\Res_{y=0}\left[(1+y)\sum_{d,n= 0}^\infty\binom{2(g-i)+n-2}{n}e_d\,y^{n+d-i-1}\right].
\end{equation}
If $i=0$ one easily sees that $m_0=e_0$. If $i\ge 1$, the evaluation of the residue gives
\begin{equation}
   m_{i}=\sum_{d=0}^i\binom{2g-i-d-2}{i-d}e_d+
\sum_{d=0}^{i-1}\binom{2g-i-d-3}{i-d-1}e_d\,,
\end{equation}
which can be rewritten as \eqref{invrel}.
\flushleft{\rm(C)}\ This is already obvious from what we have seen in this proof.
\end{proof}

\section{$D2$-$D0$ as vortices}

We  briefly recall the vortex picture of $D2$-$D0$ branes. For more on this see \cite{KY}.
Let $C\subset X$ be a smooth irreducible curve of genus $g$ which is rigid in $X$.  Suppose that a $D2$-brane is singly wrapping around $C$ and $d$ $D0$-branes are bound to it. Such a BPS system may be identified with  vortices on $C$ with magnetic flux $d$. As well-known the moduli space of such vortices is the $d$-th symmetric product of $C$ which we write as $C^{(d)}$. There exists  a  classical result by Macdonald on the Euler-Poincar{\' e} characteristics of symmetric products of smooth curves. As argued in \cite{KY}, by taking into account the fact that $D$-brane charges must be measured by Mukai vectors, it is more natural to consider a twisted version of Macdonald formula:
\begin{equation}
  \sum_{d=0}^\infty\chi(C^{(d)})y^{\pm(d+1-g)}=\iota_\pm  (y^{\inv{2}}-y^{-\inv{2}})^{2g-2}\,.
\end{equation}

A naive way to  treat $D2$-$D0$ branes in more general settings  is to extend  the vortex picture relatively and to replace $C^{(d)}$ by
$\mathcal{C}_\beta^{[d]}$. To simplify the situation considerably  we make a rather strong 
\begin{ass}\label{ass-integral}
All the fibers of $\mathcal{C}_\beta\to S_\beta$  are integral.  We have
$\mathfrak{M}_{\beta,d}=\mathcal{C}_\beta^{[d]}$ and $\mathfrak{N}_{\beta}=\bar{\mathcal{J}}_\beta$.
\end{ass}

Hence,
\begin{equation}\label{fund2}
  \sum_{d=0}^\infty\chi(\mathcal{C}^{[d]}_\beta)y^{\pm(d+1-g(\beta))}=\iota_\pm   f_\beta(y)\,.
\end{equation}
From this 
\eqref{smooth} immediately follows.

Consider for instance the case of  $\mathscr{O}_{\PP^1}(-1)\oplus \mathscr{O}_{\PP^1}(-1)$. Let $[\PP^1]$ denote the class of the $\PP^1$. Then,
\begin{equation}\label{2point}
f_{[\PP^1]}(y)=(y^{\inv{2}}-y^{-\inv{2}})^{-2}\,,
\end{equation}
and $f_{n[\PP^1]}(y)=0$ for $n>1$.  This expression can also be obtained (in a difficult way!) either by
Gromov-Witten theory \cite{FP} or by  the topological vertex formalism \cite{AKMV}.
In \S \ref{Why} we said that bubble trees in Gromov-Witten theory turn into $D0$-branes (vortices). It is  instructive to see how this happens on this simple example.
In \cite{FP}, the authors use the localization technique to evaluate the Gromov-Witten invariants. This boils down to calculate the bubbling contributions at the north and south poles of the $\PP^1$. On the other hand, the approach of the topological vertex
gives a diagram consisting  of one internal edge corresponding to the $\PP^1$  and two trivalent vertices corresponding to the north and south poles. The torus-fixed $D2$-$D0$ configurations relevant to calculate $f_{[\PP^1]}(y)$ is such that  a $D2$-brane wraps singly around the $\PP^1$ (so the $2d$ Young diagram is a single box) and $D0$-branes are localized to  the north or south poles, {\it i.e.}\ the two vertices. 

We also note that
\eqref{2point} has an interpretation as a two-point function of vertex operators \cite{Grojnowski,Nakajima} (See also \cite{KY}). This matches well with the appearance of the Schur polynomials in the approach of \cite{AKMV}.

In the simplified setting we assume later, we will  find that $f_\beta(y)$ has an 
expansion of the form
\begin{equation}\label{expansion-of-f}
  f_\beta(y)=\sum_{h=0}^{g(\beta)}n_\beta^h\, \varphi_h(y)\,,
\end{equation}
with $n_\beta^h\in \ZZ$ and $\varphi_h(y)$ being a rational function.
In fact there arise at least two natural choices of basis functions $ \varphi_h(y)$.  One choice is associated with Gopakumar-Vafa's expansion \eqref{GVexpansion} and $ \varphi_h(y)$ takes a simple form.   The other choice has a clear geometrical meaning but $ \varphi_h(y)$ takes a more complicated form.

In \cite{KY} we studied an analogous case with $X$ replaced by a projective  $K3$ surface $M$.
Corresponding to $\mathcal{C}_\beta\to S_\beta$ in the present situation we
 considered  the universal family of curves $\mathcal{C}_n\to \abs{D_n}\cong \PP^n$ where $D_n$ is  a  smooth (integral) curve of genus $n$ on  $M$.   Associated to $\mathcal{C}_n\to \abs{D_n}$  one considers  the relative compactified Jacobian $\bar{\mathcal{J}}_n \to  \abs{D_n}$ together with the Abel-Jacobi map $\mathcal{C}_n^{[d]}\to
 \bar{\mathcal{J}}_n $. 
  With the technical assumptions 
there,  all the fibers of $\mathcal{C}_n\to \abs{D_n}$ are integral and  both $\mathcal{C}_n^{[d]}$ and 
 $\bar{\mathcal{J}}_n $ are smooth. 
In \cite{KY} we proved an exact  formula for
\begin{equation}
\sum_{d=0}^\infty\chi(\mathcal{C}^{[d]}_n)y^{\pm(d+1-n)}\,.
\end{equation}

(The proof there rested on   Brill-Noether theory of sheaves on $K3$ surfaces developed by Yoshioka \cite{Yoshioka} and Markman \cite{Markman}.)

\begin{rem}
 In \cite{KKV}, the authors tried to develop an algorithm  for calculating the numbers $N_\beta^h$ and in the process of this  they also considered  $\mathcal{C}_\beta^{[d]}$. 
However in their treatment it played only an auxiliary role since their point of view is that of Gopakumar-Vafa and not of the $D2$-$D0$ picture. They applied their formalism to concrete examples and obtained satisfactory results with the caveat that  one must be careful when there are reducible fibers in the family of curves ({\em e.g.\/} the case of local $\PP^2$). In the following we will see that some of their findings can be uniformly understood once one considers the relation between the Gopakumar-Vafa picture and the   $D2$-$D0$ picture.
\end{rem}

\section{Abel-Jacobi maps and Lefschetz actions}

Let $C$ be a smooth irreducible curve of genus $g$ and let $J(C)$ be its Jacobian.
We have 
\begin{equation}
(-1)^g\Tr_{H^*(J(C))}(-1)^Hy^H=(y^{\inv{2}}-y^{-\inv{2}})^{2g}\,,
\end{equation}
 where $H$ is the image of the Cartan generator of the Lefschetz $sl(2)$ as before. Rather trivially, it  follows that 
\begin{equation}\label{CL}
\sum_{d=0}^\infty\chi(C^{(d)})y^{\pm(d+1-g)}=\iota_\pm \frac{(-1)^g\Tr_{H^*(J(C))}(-1)^Hy^H}{ (y^{\inv{2}}-y^{-\inv{2}})^{2}}\,.
\end{equation}

What  \eqref{K-GV} implies is that one may proceed relatively:
\begin{yosou}\label{CLconj}
\begin{equation}\label{CLconjeq}
\sum_{d=0}^\infty\chi(\mathcal{C}^{[d]}_\beta)y^{\pm(d+1-g(\beta))}=\iota_\pm \frac{\epsilon(\beta)\Lambda_\beta(y)}{ (y^{\inv{2}}-y^{-\inv{2}})^{2}}\,.
\end{equation}
\end{yosou}
This is not obvious at all since there can be singular fibers in $\mathcal{C}_\beta\to S_\beta$.  However, with some simplifying assumptions, this conjecture can be shown to be  true.
See Theorem \ref{mainth}. Notice that one may take 
$n_\beta^h=N_\beta^h$ and $ \varphi_h(y)=(y^{\inv{2}}-y^{-\inv{2}})^{2h-2}$
in  \eqref{expansion-of-f}.

Conjecture \ref{CLconj}  implies  relations between
$\chi(\mathcal{C}_\beta^{[d]})$ and $N_\beta^h$ again by  Proposition \ref{invprop}.
In particular, we note that \eqref{invrel} coincides with  the relation claimed in \cite{KKV}.
If $d\ge 2g-1$ the Abel-Jacobi map  $\mathcal{C}^{[d]}_\beta\to{\bar {\mathcal{J}}}_\beta$ is a fibration with fiber $\PP^{d-g}$.
Since  $m_g=N_\beta^0=\chi(\bar{\mathcal{ J}}_\beta)$, the relation  $e_d=(d+1-g)m_g$  is just a reflection of this  projective bundle structure.

 It is again instructive to see the situation for  the  $K3$ surface $M$ as before.
 There is little harm\footnote{As well-known they are birationally equivalent or  even deformation equivalent under the conditions in \cite{KY}.   See there and references therein.} in replacing $ \bar{\mathcal{J}}_n $ by the Hilbert scheme of points $M^{[n]}:=\Hilb^n_M$.  When one views $M^{[n]}$ as a compact hyperk{\" a}hler manifold\footnote{See \cite{HuybrechtsI, HuybrechtsII} for surveys. } of real dimension $4n$, 
there are a triple of K{\" a}hler forms $\omega_I,\omega_J,\omega_K$ and therefore a triple of Lefschetz actions $\rho_I,\rho_J,\rho_K:sl(2)\to \End(H^*(M^{[n]}))$. According to Verbitsky \cite{Verbitsky},
$\rho_I,\rho_J,\rho_K$ generate an action of  $so(4,1)$ on $H^*(M^{[n]})$. Gopakumar-Vafa's  $sl(2)\oplus sl(2)$  should be identified (with the replacement of $ \bar{\mathcal{J}}_n $ by $M^{[n]}$ understood)  as a subalgebra of  Verbitsky's  $so(4,1)$.

We choose one complex structure on  $M^{[n]}$ thereby viewing it as an irreducible holomorphic symplectic manifold of complex dimension $2n$.  Denote the associated  K{\" a}hler form by $\varpi$. Then  $\rho_D: sl(2)\to \End(H^*(M^{[n]}))$ should be identified  with the usual Lefschetz  action. So the lowering operator of the $sl(2)$ is represented by the wedge operation of  $\varpi$ and the raising one by its adjoint.
On the other hand,  as we will justify shortly, $\rho_L: sl(2)\to \End(H^*(M^{[n]}))$  should be identified with the holomorphic Lefschetz  action first considered by Fujiki \cite{Fujiki}.  
 So the lowering operator is represented by the wedge  operation of  a holomorphic 2-form $\sigma\in H^0(M^{[n]},\Omega_{M^{[n]}}^2)$ and the raising one by its adjoint.

Apparently  an analog of $\Lambda_\beta(y)$ is
\begin{equation}
\Lambda_n(y):=\Tr_{H^*(M^{[n]})}(-1)^{H_D}y^{H_L}\,.
\end{equation}
\begin{proposition} {\rm(Thompson's observation)}  With the convention that $\chi_y(-)$ is meant for  $\chi_{-y}(-)$ in the original sense of Hirzebruch, we have
\begin{equation}\label{Tobs}
\Lambda_n(y)=y^{-n}\chi_y(M^{[n]})\,.
\end{equation}

\end{proposition}
\begin{proof}
Since $H_D$  acts as the multiplication of  $2n-p-q$ on $H^q(M^{[n]},\Omega_{M^{[n]}}^p)$ and  we have 
$(-1)^{2n-p-q}=(-1)^{p+q}$, we see that $\Lambda_n(y)$ is  precisely  what Thompson writes as  $\mathop{\rm STr} U$ in \cite{Thompson} with $y$ being one of the eigenvalue of $U\in SL(2,\CC)$. (Amusingly, $\mathop{\rm STr} U$ is  the Rozansky-Witten invariant of a mapping torus $T_U^3$ if $U\in SL(2,\ZZ)$.)
So \eqref{Tobs} is just a consequence of  his observation.
\end{proof}

To our please we showed in  \cite{KY} that
\begin{thm}
\begin{equation}
\sum_{d=0}^\infty\chi(\mathcal{C}^{[d]}_n)y^{\pm(d+1-n)}=\iota_\pm\frac{ y^{-n}\chi_y(M^{[n]})}{(y^{\inv{2}}-y^{-\inv{2}})^{2}}\,.
\end{equation}

\end{thm}
(This was proved by comparing our  formula for $\chi(\mathcal{C}_n^{[d]})$ and  the formula for $\chi_y(M^{[n]})$ proved  earlier by G{\" o}ttsche and Soergel \cite{GS}.) Note that what corresponds to $\epsilon(\beta)$ is $(-1)^{2n}=1$.

\section{Hilbert schemes of points on nodal curves}
As a preparation for   the next section we make a digression to study Hilbert schemes of points on nodal curves. It is well-known that $C^{[d]}:=\Hilb^d_C$ coincides with $C^{(d)}$ if $C$ is smooth. This, however, is not the case when $C$ is singular.
Suppose that  $C_{g,\delta}$ is an integral curve of arithmetic genus $g$ having  $\delta$ nodes and no other singularities. In this case we say that $C_{g,\delta}$ is $\delta$-nodal. Note that $g\ge \delta\ge 0$.  Let $\nu:\tilde C_{g,\delta}\to C_{g,\delta}$ be the normalization.  Then $\chi(C_{g,\delta})=\chi(\tilde C_{g,\delta} )-\delta=2-2(g-\delta)-\delta=2+\delta-2g$. In general
\begin{equation}
  \chi(C^{(d)}_{g,\delta})=\binom{d+1+\delta-2g}{d}\,.
\end{equation}
For the Hilbert schemes we have to find the necessary  correction terms to this formula, which we now turn to.

We introduce the standard notation for multinomial coefficients.
\begin{equation}
  \binom{a_1+\cdots+a_n}{a_1,\ldots,a_n}:=\frac{(a_1+\cdots+a_n)!}{a_1!\cdots a_n!}\,,\qquad a_1\ge 0,\dots,a_n\ge 0\,.
\end{equation}

Recall that a {\em partition\/} is any sequence $\lambda=(\lambda_1,\lambda_2,\dots)$ of non-negative integers in non-increasing order and containing only finitely many non-zero terms. Then $\abs{\lambda}=\sum_i\lambda_i$ is the {\em weight\/} of $\lambda$ and we say $\lambda$ is a {\it partition of $d$\/} if $\abs{\lambda}=d$.
When $\lambda$ is a partition of $d$, we also use an alternative notation $\lambda=(1^{\delta_1}2^{\delta_2}\cdots d^{\delta_d})$ where $\delta_\ell=\#\{i\mid \lambda_i=\ell\}$ is the multiplicity of $\ell$ in $\lambda$. (Note that $\delta_\ell$ might be zero.) We set  ${\check \delta}_1=\delta_2+\cdots +\delta_d$ for convenience.

If  $\lambda=(1^{\delta_1}2^{\delta_2}\cdots d^{\delta_d})$ is a partition of $d$, we define 
$A_\lambda$ by 
\begin{equation}
 A_\lambda=1,\quad (d=0,1)\,,\qquad  A_\lambda= \prod_{\ell=2}^d (\ell-1)^{\delta_\ell},\quad (d>1)\,.
\end{equation}
\begin{proposition} 
  \begin{equation}\label{eulerhilb1}
   \chi(C^{[d]}_{g,\delta})=\sum_{\substack{\lambda=(1^{\delta_1}\cdots d^{\delta_d})\\[2pt]
 {\abs{\lambda}=d,\ \check \delta}_1\le \delta}} A_\lambda\, 
\binom{\delta}{\delta-{\check \delta}_1,\delta_2,\dots,\delta_d}
\binom{\delta_1+1+\delta-{\check \delta}_1-2g}{\delta_1}\,.
\end{equation}
\end{proposition}
\begin{proof}
We assume  $d>1$ since otherwise the assertion is trivial.

Let $\mathcal{N}$ be the zero-dimensional subscheme of the nodes on  $C_{g,\delta}$ so that its length is $\# \mathcal{N}=\delta$.
For a given partition  $\lambda=(1^{\delta_1}\cdots d^{\delta_d})$ such that $\delta\ge {\check \delta}_1$ we partition
$\mathcal{N}$ into mutually disjoint union
$\mathcal{N}=\coprod_{\ell=1}^d  \mathcal{N}_\ell^\lambda$
with
$\#\mathcal{N}_1^\lambda=\delta-{\check \delta}_1$
and $\#\mathcal{N}_\ell^\lambda=\delta_\ell$ $(\ell >1)$.  
In $C^{(d)}_{g,\delta}$ we consider configurations where  a multiplicity $\ell$ point collides  with each node of $\mathcal{N}_\ell^\lambda$ for each $\ell>1$.
Then $C^{[d]}_{g,\delta}$
 differs from $C^{(d)}_{g,\delta}$ in that at each node of $\mathcal{N}_\ell^\lambda$
 a multiplicity $\ell$ point is replaced by the punctual Hilbert scheme of length $\ell$.

After  some mental exercise,  $\chi(C^{[d]}_{g,\delta})$ is then found to be
\begin{equation}
\sum_{\substack{\lambda=(1^{\delta_1}\cdots d^{\delta_d})\\[2pt]
 {\abs{\lambda}=d,\ \check \delta}_1\le \delta}}\, \sum_{\mathcal{N}=\coprod_{\ell=1}^d  \mathcal{N}_\ell^\lambda}\chi\left(\bigl(C_{g,\delta}\setminus{\textstyle \coprod}_{\ell=2}^d  \mathcal{N}_\ell^\lambda\bigr)^{(\delta_1)}\right)\prod_{\ell=2}^d(\chi(H_\ell)-1)^{\delta_\ell}\,,
\end{equation}
where $H_\ell$ is the punctual Hilbert scheme of length $\ell$ supported at a node. 
To proceed we have to calculate $\chi(H_\ell)$.
According to Ran  \cite{Ran},   $H_\ell$ is  a rational chain
\begin{equation}
H_\ell\cong R_1\mathop{\cup}\limits_{p_1}R_2 \mathop{\cup}\limits_{p_2}\cdots \mathop{\cup}\limits_{p_{\ell-2}} R_{\ell-1}\,,
\end{equation}
where $R_1,\dots,R_{\ell-1}$ are $\PP^1$'s  and  $p_1,\dots,p_{\ell-2}$ are nodes.
Therefore    $\chi(H_\ell)=(\ell-1)\times 2-(\ell-2)=\ell$ and we understand why  the factor $A_\lambda$ arises.
 
Next we easily  see that
\begin{equation}
 \chi((C_{g,\delta}\setminus{\textstyle \coprod}_{\ell=2}^d  \mathcal{N}_\ell^\lambda)^{(\delta_1)})= \binom{\delta_1+1+\delta-{\check \delta}_1-2g}{\delta_1}\,.  
\end{equation}

Finally,  it should be noticed that  there are 
$
  \binom{\delta}{\delta-{\check \delta}_1,\delta_2,\dots,\delta_d}
$
 ways to partition $\mathcal{N}$  into $\coprod_{\ell=1}^d \mathcal{N}_\ell^\lambda$.
\end{proof}

\Yboxdim 3.2pt
The above formula of $\chi(C^{[d]}_{g,\delta})$ is rather complicated and not that convenient in practice. Actually there exists a neat formula for the generating function  as we explain below.
However, before going to that,  it will be worthwhile (and fun!)  to look at  several examples of this formula: 
  \begin{equation*}
\begin{array}{ccccccc}
\chi(C_{g,\delta})&=&2+\delta-2g&&&&\\
&&\yng(1)&&&&\\[2mm]
 \chi(C^{[2]}_{g,\delta})&=&\binom{3+\delta-2g}{2}&+&\delta&&\\
&&\yng(1,1)&  &\yng(2)&&\\[2mm]
\chi(C^{[3]}_{g,\delta})&=&\binom{4+\delta-2g}{3}&+&\delta(1+\delta-2g)&+&2\delta\\
&&\yng(1,1,1)&  &\yng(2,1)&&\yng(3)
\end{array}
\end{equation*}
These agree with the sample computations of  \cite{KKV}.
We can give further examples:
\begin{equation*}
\begin{array}{ccccccc}
\chi(C^{[4]}_{g,\delta})
&=&
\binom{5+\delta-2g}{4}
&+&
\delta\binom{2+\delta-2g}{2}
&+&
\binom{\delta}{2}\\[1mm]
&&\yng(1,1,1,1)&  &\yng(2,1,1)&&\yng(2,2)\\[2mm]
&+&
2\delta(1+\delta-2g)
&+&
3\delta&&\\
&&\yng(3,1)&  &\yng(4)&&
\end{array}
\end{equation*}
\begin{equation*}
\begin{array}{ccccccc}
\chi(C^{[5]}_{g,\delta})&=&
\binom{6+\delta-2g}{5}
&+&
\delta\binom{3+\delta-2g}{3}
&+&
\binom{\delta}{2}(\delta-2g)\\[1mm]
&&\yng(1,1,1,1,1)&  &\yng(2,1,1,1)&&\yng(2,2,1)\\[2mm]
&+&2\delta\binom{2+\delta-2g}{2}
 &+&
2\delta(\delta-1)
&+&
3\delta(1+\delta-2g)\\[1mm]
&&\yng(3,1,1)&  &\yng(3,2)&&\yng(4,1)\\[2mm]
&+&4\delta&&&&\\
&&\yng(5)&  &&&
\end{array}
\end{equation*}
\begin{equation*}
\begin{array}{ccccccc}
\chi(C^{[6]}_{g,\delta})&=&
\binom{7+\delta-2\,g}{6}
&+&
\delta{4+\delta-2\,g\choose 4}
&+&
{\delta\choose 2}{1+\delta-2\,g\choose 2} \\[2mm]
&&\yng(1,1,1,1,1,1)&  &\yng(2,1,1,1,1)&&\yng(2,2,1,1)\\[3mm]
&+&
{\delta\choose 3}
&+&
2\,\delta{3+\delta-2\,g\choose 3}
&+&
2\,\delta \left( \delta-1 \right) 
 \left( \delta-2\,g \right) \\[1mm]
&&\yng(2,2,2)&  &\yng(3,1,1,1)&&\yng(3,2,1)\\[3mm]
&+&
2^2\binom{\delta}{2}
&+&
3\,\delta{2+\delta-2\,g\choose 
2}
&+&
3\,\delta \left( \delta-1 \right)\\[1mm]
&&\yng(3,3)&  &\yng(4,1,1)&&\yng(4,2)\\[3mm]
&+&
4\,\delta \left( 1+\delta-2\,g \right)
&+&
5\,\delta & & \\
&&\yng(5,1)&  &\yng(6)&&\\[2mm]
\end{array}
\end{equation*}

Let us now turn to the generating function:
\begin{proposition}\label{nodalMacdonald}
  \begin{equation}\label{eulerhilb2}
\sum_{d=0}^\infty\chi(C^{[d]}_{g,\delta})y^{\pm(d+1-g)}=\iota_\pm\, (y^{\inv{2}}-y^{-\inv{2}})^{2g-2}\left(1+ (y^{\inv{2}}-y^{-\inv{2}})^{-2}\right)^\delta\,.
\end{equation}
\end{proposition}

\begin{proof}
 By eliminating the sum over $\delta_1$ in \eqref{eulerhilb1} one obtains that
\begin{equation}
  \begin{split}
      \chi(C^{[d]}_{g,\delta})=&
\sum_{\substack{\delta_2\ge 0,\dots,\delta_d\ge 0\\[2pt]
   \sum_{\ell \ge 2} \ell \delta_\ell \le d\\ \sum_{\ell \ge 2} \delta_\ell\le \delta}}
\binom{\delta}{\sum_{\ell \ge 2} \delta_\ell}\binom{\sum_{\ell \ge 2} \delta_\ell}{\delta_2,\dots,\delta_d}\left(\prod_{\ell=2}^d(\ell-1)^{\delta_\ell}\right)
\\
&\qquad\quad\quad \times\binom{d-\sum_{\ell \ge 2} \ell \delta_\ell+1+\delta- \sum_{\ell \ge 2} \delta_\ell-2g}{d-\sum_{\ell \ge 2} \ell \delta_\ell }\,.
\end{split}
\end{equation}
Then consideration of the generating function leads to
\begin{equation}
 \begin{split}
     \sum_{d=0}^\infty\chi(C^{[d]}_{g,\delta})y^d&=
\sum_{k=0}^\delta\,\,\sum_{\substack{\delta_2\ge 0,\delta_3\ge 0,\dots\\[2pt]
 k= \sum_{\ell \ge 2} \delta_\ell}}\binom{\delta}{k}\binom{k}{\delta_2,\delta_3,\dots}\\
&\quad\times\left(\prod_{\ell\ge 2}(\ell-1)^{\delta_\ell}\right)
y^{\sum_{\ell \ge 2} \ell \delta_\ell}(1-y)^{2g-\delta+k-2}\,,
\end{split}
\end{equation}
where  we understand the sequence $\delta_2,\delta_3,\dots$ contains only finitely many non-zero terms.
The right hand side can be rewritten as
\begin{equation}
 \begin{split}
&(1-y)^{2g-\delta-2}\sum_{k=0}^\delta\binom{\delta}{k}\\
&\quad\times \sum_{\substack{\delta_2\ge 0,\delta_3\ge 0,\dots\\[2pt]
 k= \sum_{\ell \ge 2} \delta_\ell}}
\binom{k}{\delta_2,\delta_3,\dots}
\prod_{\ell\ge 2}\left\{(\ell-1)y^\ell(1-y)\right\}^{\delta_\ell}\,.
\end{split}
\end{equation}
The multinomial theorem simplifies this as
\begin{equation}
(1-y)^{2g-\delta-2}\sum_{k=0}^\delta\binom{\delta}{k}\left(\sum_{\ell \ge 2}(\ell-1)y^\ell(1-y)\right)^k\,.
\end{equation}
By summing up the geometric series one obtains
\begin{equation}
  (1-y)^{2g-\delta-2}\sum_{k=0}^\delta\binom{\delta}{k}\left(\frac{y^2}{1-y}\right)^k\,.
\end{equation}
By the binomial theorem this is equal to 
\begin{equation}
(1-y)^{2g-\delta-2}\left(1+\frac{y^2}{1-y}\right)^\delta\,.
\end{equation}
It thus follows that
\begin{equation}\label{en}
     \sum_{d=0}^\infty\chi(C^{[d]}_{g,\delta})y^d=(1-y)^{2g-2}\left(1+\frac{y}{(1-y)^2}\right)^\delta\,.
\end{equation}
To see that this leads to \eqref{eulerhilb2} is easy.
\end{proof}

 Let $J(C_{g,\delta})$ be the generalized Jacobian of $C_{g,\delta}$ and $\overline{J(C_{g,\delta})}$ its compactification.
\begin{proposition}\label{CJformulaprop}
\begin{equation}\label{CJformula}
(-1)^g\Tr_{H^*(\overline{J(C_{g,\delta})})}(-1)^Hy^H=(y^{\inv{2}}-y^{-\inv{2}})^{2(g-\delta)}\left((y^{\inv{2}}-y^{-\inv{2}})^{2}+1\right)^\delta\,.
\end{equation}
\end{proposition}
\begin{proof}
This is (with a correction of sign)  implicit in \cite{KKV}. (See the discussion around eq.~(5.8) of \cite{KKV}.)
We have  an exact sequence of commutative algebraic groups\footnote{A line bundle on   $C_{g,\delta}$  can be  obtained from one on  $\tilde{C}_{g,\delta}$ by gluing  the fibers over those points on $\tilde{C}_{g,\delta}$  which are  identified  to form nodes on $C_{g,\delta}$.
For each node the ways to glue two lines are parametrized by  $GL(1,\CC)=\CC^\times$.}
\begin{equation}
1\to (\CC^\times)^\delta\to J(C_{g,\delta})\xrightarrow{\nu^*} J(\tilde{C}_{g,\delta})\to 1\,.
\end{equation}
Hence $\overline{J(C_{g,\delta})}$ is the product of   $J(\tilde{C}_{g,\delta})$ and 
 the compactification of  $(\CC^\times)^\delta$. The latter  is  given by
\begin{equation}
(\PP^1)^\delta/\sim\ \cong\  (C_{\ast})^\delta\,,
\end{equation}
where $\sim$ is a certain  equivalence relation essentially identifying $0$ and $\infty$ of each $\PP^1$ and $C_\ast$ is a rational curve with a single node. See IIIb, \S 5 in \cite{Mumford} for  an exposition. Since $C_\ast$  is obtained by pinching  a homologically non-trivial 1-cycle from an elliptic curve,
we obtain
\begin{equation}
(-1)\Tr_{H^*(C_{\ast})}(-1)^Hy^H=(y^{\inv{2}}-y^{-\inv{2}})^{2}+1\,.
\end{equation}
Combining this with 
\begin{equation}
(-1)^{g-\delta}\Tr_{H^*(J(\tilde{C}_{g,\delta}))}(-1)^Hy^H=(y^{\inv{2}}-y^{-\inv{2}})^{2(g-\delta)}\,,
\end{equation}
we easily obtain \eqref{CJformula}.
\end{proof}

The following may be viewed as an extension of \eqref{CL}.
\begin{thm}\label{nodalCL}
\begin{equation}
\sum_{d=0}^\infty\chi(C^{[d]}_{g,\delta})y^{\pm(d+1-g)}=\iota_\pm\, \frac{(-1)^g\Tr_{H^*(\overline{J(C_{g,\delta})})}(-1)^Hy^H}{ (y^{\inv{2}}-y^{-\inv{2}})^{2}}\,.
\end{equation}
\end{thm}
\begin{proof} An immediate consequence of Proposition \ref{nodalMacdonald} and 
Proposition \ref{CJformulaprop}.
\end{proof}
\section{Severi varieties}
We now  investigate how the fibration  structure of  $\mathcal{C}_\beta\to S_\beta$
is directly reflected to   properties  of $f_\beta(y)$. We start by
\begin{definition}  The Severi variety of $\delta$-nodal curves is defined by
\begin{equation} 
V_{\beta,\delta}=\{ x\in S_\beta \mid \text{\rm $(\mathcal{C}_\beta)_x$ has $\delta$ nodes and no other singularities}\}\,,
\end{equation}
where $(\mathcal{C}_\beta)_x$ is the fiber over $x$.

\end{definition}

Severi varieties were originally studied by Severi (and corrected by Harris) for curves of fixed degrees in $\PP^2$. There are numerous works on  Severi varieties of other surfaces. See for instance \cite{CS, FlaminiI}.  Here we are extending the definition for  Calabi-Yau threefolds assuming  an appropriate family of curves $\mathcal{C}_\beta\to S_\beta$. For attempts in defining Severi varieties of  threefolds see \cite{FlaminiII,FlaminiIII} and references therein.
One of the principal issues in the studies of Severi varieties is  whether or not they are regular.
The standard way to investigate this is to resort to a deformation theory.  There are many works for surfaces and these show that in most cases  Severi varieties of surfaces are  regular  but  a care is needed when they are of general type. See again \cite{CS, FlaminiI}.
 If $V_{\beta,\delta}$ is regular,
the $\delta$ nodes can be independently smoothed and $\dim V_{\beta,\delta}=\dim S_\beta-\delta$. Below we assume the regularity of $V_{\beta,\delta}$ for simplicity so that $\chi(V_{\beta,\delta})$ is meaningful.

In order to appreciate  the roles of the Severi varieties we wish to add one more strong 
\begin{ass}
All the singular fibers of $\mathcal{C}_\beta\to S_\beta$ are nodal. Hence,
$S_\beta=\coprod_{\delta=0}^{g(\beta)}V_{\beta,\delta}$.
\end{ass}
In this ideal situation, 
\begin{proposition}
\begin{equation}\label{nodalMacdonald2}
  \begin{split}
  &\sum_{d=0}^\infty\chi(\mathcal{C}^{[d]}_\beta)y^{\pm(d+1-g(\beta))}\\
&\ \ =\iota_\pm \sum_{\delta=0}^{g(\beta)}\chi(V_{\beta,\delta})\,  (y^{\inv{2}}-y^{-\inv{2}})^{2g(\beta)-2}\left(1+(y^{\inv{2}}-y^{-\inv{2}})^{-2}\right)^\delta\,.
\end{split}
\end{equation}
 \end{proposition} 
 \begin{proof} 
 By our assumptions, we see (as in \cite{YZ})
\begin{equation}\label{tmp1}
\sum_{d=0}^\infty\chi(\mathcal{C}^{[d]}_\beta)y^{\pm(d+1-g(\beta))}=\sum_{\delta=0}^{g(\beta)}\chi(V_{\beta,\delta})\,  \sum_{d=0}^\infty\chi(C^{[d]}_{g(\beta),\delta})y^{\pm(d+1-g(\beta))}\,,
\end{equation}

where $C_{g(\beta),\delta}$ is any integral $\delta$-nodal curve of arithmetic genus $g(\beta)$. 
Then
\eqref{nodalMacdonald2} follows immediately from Proposition \ref{nodalMacdonald}.
\end{proof}
Notice that
\eqref{nodalMacdonald2} gives  another expansion of the form \eqref{expansion-of-f}.
We have now reached   the main assertion:
\begin{thm}\label{mainth}
Under our assumptions Conjecture \ref{CLconj} is true.
\end{thm}
\begin{proof}
Observe that
\begin{equation}\label{tmp2}
\begin{split}
\epsilon(\beta)\Lambda_\beta(y)&=\epsilon(\beta)\sum_{\delta=0}^{g(\beta)}(-1)^{\dim S_\beta}\chi(V_{\beta,\delta})
\Tr_{H^*(\overline{J(C_{g(\beta),\delta})})}(-1)^Hy^H\\
&=\sum_{\delta=0}^{g(\beta)}\chi(V_{\beta,\delta})(-1)^{g(\beta)}
\Tr_{H^*(\overline{J(C_{g(\beta),\delta})})}(-1)^Hy^H\,,
\end{split}
\end{equation}
where $C_{g(\beta),\delta}$ is again any integral $\delta$-nodal curve of arithmetic genus $g(\beta)$. 
Use  \eqref{tmp1}, \eqref{tmp2}  and  Theorem \ref{nodalCL}.
 \end{proof}
 
The relations between  $\chi(V_{\beta,\delta})$ and $N_\beta^h$ are as follows:
\begin{proposition} 
\begin{align} 
  N_{\beta}^{g(\beta)-i}&=\sum_{\delta=i}^{g(\beta)} \binom{\delta}{i}\chi(V_{\beta,\delta})\,,\label{VtoN}\\
  \chi(V_{\beta,\delta})&=\sum_{i=\delta}^{g(\beta)}(-1)^{i+\delta}\binom{i}{\delta}N_{\beta}^{g(\beta)-i}\,. \label{NtoV}
\end{align}
In particular, we have
\begin{equation}\label{VtoNspecial}
N_{\beta}^{g(\beta)}=\sum_{\delta=0}^{g(\beta)}\chi(V_{\beta,\delta})=\chi(S_\beta)\,,\qquad
N_{\beta}^{0}=\chi(V_{\beta,g(\beta)})=\chi(\bar{{\mathcal{J}}}_\beta)\,.
\end{equation}
\end{proposition}
\begin{proof}
To prove \eqref{VtoN} it suffices to binomially expand $(\cdots)^\delta$ on the right hand side of \eqref{nodalMacdonald2} and then
notice
\begin{equation}
  \sum_{\delta=0}^{g(\beta)}\sum_{i=0}^\delta=
  \sum_{i=0}^{g(\beta)}\sum_{\delta=i}^{g(\beta)}\,.
\end{equation}
The inversion relation that leads to \eqref{NtoV} is well-known. The last equality in \eqref{VtoNspecial} is due to a Yau-Zaslow type argument \cite{YZ}.
\end{proof}

The way Gopakumar and Vafa introduced $N_\beta^h$
was representation theoretic and was driven by the motivation of finding a convenient expansion basis as in \eqref{GVexpansion}.  However, to the best of my understanding, the direct geometrical characterization of $N_\beta^h$ has not yet been available. 
The relation \eqref{VtoN} seems to give  an intuitive feeling about the geometrical meaning of $N_\beta^{g(\beta)-i}$.  In general, one can produce a node on  a curve   by pinching a handle or gluing two points. Conversely given a $\delta$-nodal curve of arithmetic genus $g(\beta)$ one can make a smooth curve of genus $g(\beta)-i$,  where $i\leq \delta$,  by ungluing (or partially normalizing) $i$ nodes and unpinching $\delta-i$ nodes. Of course  there are $\binom{\delta}{i}$ ways to choose such $i$ nodes from the  entire nodes of the curve.

\begin{rem}
If $V_{\beta,\delta}=\emptyset$ for all $\delta>i$ then $N_\beta^{g(\beta)-i}=\chi(V_{\beta,i})$. This was the claim in \cite{KKV}. However,  such an assumption  depends on $i$ and cannot be satisfied for all $i$ simultaneously unless everything is empty. In general the relation between $N_\beta^h$ and $\chi(V_{\beta,\delta})$ has to be as above.
\end{rem}

To summarize, in our extremely ideal situation we can declare any of
\begin{equation}
\{\chi(\mathcal{C}_\beta^{[d]})\}_{d=0}^{g(\beta)}\,, \quad \{N_\beta^h\}_{h=0}^{g(\beta)}\,, \quad 
\{ \chi(V_{\beta,\delta})\}_{\delta=0}^{g(\beta)}\,,
\end{equation}
 as a  {\em fundamental set of  invariants}. They can be converted from one to another.

\end{document}